\documentclass[aps,preprint]{revtex4}%
\usepackage{amsfonts}
\usepackage{amsmath}
\usepackage{amssymb}
\usepackage[dvipdfmx]{graphicx}
\usepackage{braket}
\usepackage{amsmath}
\usepackage{amsfonts}
\usepackage{amssymb}
\usepackage{color}%
\providecommand{\U}[1]{\protect\rule{.1in}{.1in}}
\providecommand{\U}[1]{\protect\rule{.1in}{.1in}}
\begin{document}
\title[Configuration Interaction with Antisymmetrized Geminal Powers]{\textbf{Configuration Interaction with Antisymmetrized Geminal Powers}}
\author{Wataru Uemura, Shusuke Kasamatsu and Osamu Sugino}
\affiliation{The Institute for Solid State Physics, The University of Tokyo, 5-1-5
Kashiwanoha, Kashiwa, Chiba, 277-8581, Japan}
\keywords{many-body wavefunction, configuration interaction, antisymmetrized geminal power}
\pacs{PACS number}

\begin{abstract}
To avoid the combinatorial computational cost of configuration interaction
(CI), we have previously introduced the symmetric tensor decomposition CI
(STD-CI) method, where we take advantage of the antisymmetric nature of the
electronic wave function and express the CI coefficients compactly as a series
of Kronecker product states (STD series) [W. Uemura and O. Sugino, Phys. Rev.
Lett. 109, 253001 (2012)]. Here we extend the variational degrees of freedom
by using different molecular orbitals for different terms in the STD series.
This scheme is equivalent to the linear combination of the
Hartree-Fock-Bogoliubov state or the antisymmetrized geminal powers (AGP). The
total energy converges very rapidly within 0.72 $\mu$Hartree taking only 10
terms for the water molecule, and the convergence is likewise fast for Hubbard
tetramers. The computational cost scales as the fifth power of the number of
electrons and the square of the number of terms in the STD series, indicating the
promise of this AGP-based scheme for highly accurate and efficient computation
of quantum systems.

\end{abstract}
\maketitle

\section{Introduction}

Determining the ground state of a many-body system is the most basic problem
in many fields of science, such as condensed matter physics, nuclear physics
and quantum chemistry. Many numerical approaches have been developed so far to
obtain the wavefunction. In the configuration interaction (CI)\cite{shavitt},
multiple quasiparticle states are generated from a mean-field wavefunction and
the coefficients for their linear combination are determined by solving an
eigenvalue problem, while in the variational Monte Carlo (VMC)\cite{foulkes},
the quasiparticles are generated stochastically to obtain the expectation
value of the total energy to be variationally determined. In the tensor
network (TN) framework\cite{springer}, the total energy is variationally
determined using specific quasiparticle states generated according to an
assumed TN. In those popular schemes, accessible degrees of freedom are
usually not very large. This is particularly the case for CI albeit being the
most versatile in that CI is free from the well-known negative sign problem of
QMC and the dimensional restriction of the tensor network\cite{springer}. In
this view, extending the degrees of freedom accessible by CI is a very
important problem.

In the field of quantum chemistry, the CI wavefunction of an $N$-electron
system is expanded as
\begin{equation}
\Psi\left(  x_{1}\cdots x_{N}\right)  =\sum_{i_{1}\cdots i_{N}=1}^{M}%
A_{i_{1}\cdots i_{N}}\psi_{i_{1}}\left(  x_{1}\right)  \cdots\psi_{i_{N}%
}\left(  x_{N}\right)  , \label{CIMO}%
\end{equation}
in terms of the molecular orbitals (MOs) $\psi_{i}\left(  x\right)  $, which
are represented as a linear combination of the atomic orbitals (AOs) $\phi
_{a}$, as
\begin{equation}
\psi_{i}\left(  x\right)  =\sum_{a=1}^{M}U_{ia}\phi_{a}\left(  x\right)  ,
\label{MOAO}%
\end{equation}
using elements of an orthogonal matrix $U$. The CI coefficients $A_{i_{1}%
\cdots i_{N}}$, which are elements of an antisymmetric tensor of order
$N$\cite{rank} and dimension $M$, are varied to minimize the total energy and
thus the full-CI calculation is, in practice, hampered by the degrees of
freedom that grow combinatorially with $N$. \color{black} Usually, the CI
series is truncated to make the computation tractable; however, when starting
from the Hartree-Fock (HF) wavefunction by taking the permutation tensor
$\epsilon_{i_{1}\cdots i_{N}}$ in place of $A_{i_{1}\cdots i_{N}}$ and the
Hartree-Fock (HF) orbital $\psi_{i}^{HF}$ in place of $\psi_{i}$, the CI
series usually converges very slowly\cite{shavitt}. The convergence does not,
in general, speed up drastically even when starting from multiple Slater
determinants. Instead, use of localized MOs in place of the canonical HF
orbitals is known to be effective for this purpose\cite{LMO}, and, recently,
significant speed-up was achieved by using non-orthonormal Slater determinants
comprised of non-orthonormal MOs that augment the HF orbitals\cite{Goto}. It
is therefore expected that CI can be made more practicable by using optimal
MOs for this purpose.

Apart from CI, the many-body perturbation theory such as the coupled cluster
(CC)\cite{bartlett} has been developed to collect infinite correction terms,
thereby enabling the correction of the total energy of the water molecule, for
example, within the error of 0.53 mHartree by taking up to triple excitations
(CCSDT). However, the computational cost scales as $O(N^{8})$, which is still
prohibitively high for application to many systems of interest. Thus,
accelerating the convergence of the CI series should open up the possibility
to treat, with unprecedented accuracy, a whole new class of problems in
electronic structure theory of strongly correlated systems.

The key to the development that we present here is in a compact representation
of the antisymmetric tensor $A_{i_{1}\cdots i_{N}}$. With this in mind, we
proposed in our previous work\cite{uemura2012} to represent $A_{i_{1}\cdots
i_{N}}$ as a product of $\epsilon_{i_{1}\cdots i_{N}}$ and a symmetric tensor
$S_{i_{1}\cdots i_{N}}$ and then to decompose the latter into a series of
Kronecker products of a set of vectors such that%
\begin{equation}
S_{i_{1}\cdots i_{N}}=\sum_{r=1}^{K}c_{i_{1}}^{r}\cdots c_{i_{N}}^{r}\text{.}
\label{STD}%
\end{equation}
Such decomposition of an order-$N$ tensor into a linear combination of rank-$1$ tensors is known in the literature as the canonical
decomposition\cite{CANDECOMP} or the parallel factor
decomposition\cite{PARAFAC}. In our approach, which we call the symmetric
tensor decomposition CI (STD-CI), variational parameters are the set of
vectors $c$ and the orthogonal matrix $U$ and thus the degrees of freedom are
$KM+M\left(  M-1\right)  /2$. The number of order-$1$ tensors (i.e., the rank
of the tensor decomposition $K$\cite{rank}), was found to be relatively few
for small molecules (H$_{2}$, He$_{2}$, and LiH) and the Hubbard tetramer
(FIG. 1-4). \ This suggests effectiveness of the tensor decomposition in
compactly describing the wavefunction with a rapidly converging series (STD
series) when the molecular orbitals are optimized together. Note that each
term in the STD series contains all possible Slater determinants, which means
that all the terms of the CI series are regrouped differently to form
different terms of the STD series. The computational cost thereby required is
proportional to $K^{2}M^{6}$ when using the algorithm to handle the
permutation tensor developed in Ref.\cite{uemura2012}, so that the
effectiveness of STD-CI depends on the the rank of the decomposition $K$
required to accurately express the wavefunction. When applied to larger
systems, however, we have found that the rank $K$ needed for sufficient
convergence is increased in general. In addition, the computation suffers from
loss of significance in the floating-point arithmetic as will be shown below.
Therefore, an improvement in the tensor decomposition is clearly desirable.

In this context, one may use a more elaborate tensor decomposition, such as
the tensor train network (TTN) suitable for one-dimensional
systems\cite{white1992, white1993} or the projected entangled pair states
(PEPS) extended to two-dimensional systems\cite{PEPS}. Indeed, the former
approach was applied to several molecular systems\cite{chan2002}. However,
here we take a different approach: within the canonical decomposition
technique, the degrees of freedom are extended by using different
non-orthonormal MO coefficients for different terms in the STD series. That
is, we use a general complex matrix $U_{ia}^{r}$ for the MO coefficient
although it was described as $c_{i}^{r}U_{ia}$ in the original STD-CI scheme,
extending thereby the degrees of freedom to $KM^{2}$. We will call our new
scheme the extended STD (ESTD). As will be shown below, each term in the STD
series corresponds to a different Hartree-Fock-Bogoliubov (HFB)
wavefunction\cite{HFB}. Since the HFB wavefunction is called alternatively the
antisymmetrized geminal power (AGP)\cite{coleman1, coleman2}, ESTD may also be
referred to as a linear combination of AGP.

The AGP wavefunction has been used for variational calculations in a different
context. In quantum chemistry, a single AGP (i.e., $K=1$) has been used as the
trial wavefunction to obtain the potential energy surface of small
molecules\cite{Ortiz1981, Straroverov2002, Scuseria2011, Kobayashi2014}. In
variational Monte Carlo (VMC) calculations, single AGP\cite{weiner, eric,
sorella, Tahara2008} or multi-AGP\cite{bajdich2006, bajdich2008} is multiplied
by a correlation factor of the Jastrow type to form a Jastrow-AGP (JAGP) trial
wavefunction; the optimization is done only for the Jastrow parameters with
the MO\ coefficients kept unchanged from the initial HF orbitals or the
initial natural orbitals constructed by performing a small preparatory CI
calculation. In nuclear physics, multi-quasiparticle states are generated from
a single AGP and are linearly combined to describe the wavefunction. This
computational scheme is called the generator coordinate method (GCM)\cite{GCM}%
. The total-energy scheme of GCM was formulated by Onishi and
Yoshida\cite{Onishi1966}, and the mathematical structure was studied recently
by other groups\cite{doba, Mizusaki2012}. Contrary to those earlier works, we
fully optimize the MO coefficients in our ESTD scheme without introducing the
correlation factor. In the present scheme, we use an algorithm to decompose
the permutation tensor into products of the second order permutation tensors
allowing thereby to reduce the computational scaling to $K^{2}M^{5}$. Contrary
to STD-CI, this scheme does not require orthonormalization of AOs so that one
can take advantage of the AO basis that is spatially localized and chemically
comprehensible. We will show below that the STD series can be
drastically shortened for the H$_{\text{2}}$O molecule and Hubbard tetramers.
This may open up the possibility to interpret strongly correlated systems in
terms of MOs. In the next section, we introduce the outline of our formalism.
We restrict our study to systems with even number of electrons throughout this paper.

\section{Formalism}

In the original STD-CI, we have applied the canonical decomposition to the
symmetric part of the CI\ coefficient (Eq.\ (\ref{STD})), so that the
wavefunction Eq.\ (\ref{CIMO}) is rewritten using Eq.\ (\ref{MOAO}) as%
\begin{equation}
\Psi\left(  x_{1}\cdots x_{N}\right)  =\sum_{a_{1}\cdots a_{N}=1}^{M}%
A_{a_{1}\cdots a_{N}}\phi_{a_{1}}\left(  x_{1}\right)  \cdots\phi
_{a_{N}}\left(  x_{N}\right)
\end{equation}
with the CI coefficient becoming
\begin{equation}
A_{a_{1}\cdots a_{N}}=\sum_{i_{1}\cdots i_{N}=1}^{M}\sum_{r=1}^{K}c_{i_{1}%
}^{r}\cdots c_{i_{N}}^{r}\epsilon_{i_{1}\cdots i_{N}}U_{i_{1}a_{1}}\cdots
U_{i_{N}a_{N}}, \label{decomposition}%
\end{equation}
where $K$ is the rank of the symmetric tensor decomposition and $c_{i}^{r}$ is
a set of complex vectors. We will take the convention for the subscript such
that $i,j,\ldots$ correspond to MOs and $a,b,\ldots$ to AOs. In the extended
STD-CI (ESTD), we expand the degrees of freedom by replacing the product
$c_{i}^{r}U_{ia}$ with a general complex matrix $U_{i\alpha}^{r}$ such that
\begin{equation}
A_{a_{1}\cdots a_{N}}=\sum_{r=1}^{K}\sum_{i_{1}\cdots i_{N}=1}^{M}%
\epsilon_{i_{1}\cdots i_{N}}U_{i_{1}a_{1}}^{r}\cdots U_{i_{N}a_{N}}^{r}%
\equiv\sum_{r=1}^{K}A_{a_{1}\cdots a_{N}}^{r}. \label{extended}%
\end{equation}
Using the fact that the permutation tensor of order $N$ for even $N$ can be
decomposed into a product of the second order permutation tensors as%
\begin{align}
\epsilon_{i_{1}\cdots i_{N}}  &  =\frac{1}{\left(  N/2\right)  !2^{N/2}}%
\sum_{\sigma\in S_{N}}\mathrm{sgn}\left(  \sigma\right)  \epsilon
_{i_{\sigma\left(  1\right)  }i_{\sigma\left(  2\right)  }}\cdots
\epsilon_{i_{\sigma\left(  N-1\right)  }i_{\sigma\left(  N\right)  }}\\
&  =\frac{1}{\left(  N/2\right)  !2^{N/2}}\hat{A}(\epsilon_{i_{1}i_{2}%
}\epsilon_{i_{3}i_{4}}\cdots\epsilon_{i_{N-1}i_{N}}),
\end{align}
where $S_{N}$ is the symmetric group of degree $N$ and $\hat{A}$ is the
antisymmetrizer, Eq.\ (\ref{extended}) can be represented as%
\begin{equation}
A_{a_{1}\cdots a_{N}}^{r}=\frac{1}{\left(  N/2\right)  !2^{N/2}}\hat{A}%
(\gamma_{a_{1}a_{2}}^{r}\gamma_{a_{3}a_{4}}^{r}\cdots\gamma_{a_{N-1}a_{N}}%
^{r})
\end{equation}
using the antisymmetrized geminal
\begin{equation}
\gamma_{ab}^{r}\equiv\sum_{ij}\epsilon_{ij}U_{ia}^{r}U_{jb}^{r}. \label{gamma}%
\end{equation}
This shows that our ESTD is an extension of the antisymmetrized geminal power
(AGP) \cite{coleman1, coleman2} in that AGPs are linearly combined. In the
second-quantization form, the state vector is given by
\begin{equation}
\left\vert \Psi\right\rangle =\sum_{r=1}^{K}\left\vert \gamma^{r}%
\right\rangle
\end{equation}
with
\begin{align}
\left\vert \gamma^{r}\right\rangle  &  =\frac{1}{\left(  N/2\right)  !2^{N/2}%
}\left[  \sum_{ab=1}^{M}\gamma_{ab}^{r}c_{a}^{\dagger}c_{b}^{\dagger}\right]
^{N/2}\left\vert 0\right\rangle \\
&  =\left.  \exp\left[  \frac{t}{2}\sum_{ab}\gamma_{ab}^{r}c_{a}^{\dagger
}c_{b}^{\dagger}\right]  \left\vert 0\right\rangle \right\vert _{t^{N/2}},
\label{ESTDexp}%
\end{align}
where the subscript $t^{N/2}$ means taking a coefficient of degree $N/2$ from
the Hartree-Fock-Bogoliubov state\cite{HFB}. This indicates a formal
similarity of the ESTD wavefunction with that of the generator coordinate
method (GCM)\cite{GCM}, and thus one can take advantage of the formulas
developed for GCM. For example, we can use the matrix elements derived by
Onishi and Yoshida\cite{Onishi1966} such as%
\begin{align}
\left\langle \gamma^{r_{2}}\right.  \left\vert \gamma^{r_{1}}\right\rangle  &
=\,\left.  \exp\left[  \frac{1}{2}\mathrm{tr}\,\mathrm{log}\left(
1+\gamma^{r_{2}\dagger}\gamma^{r_{1}}t\right)  \right]  \right\vert _{t^{N/2}%
}\label{GCM1}\\
&  \equiv\left.  \mathrm{pf}\left(  1+\gamma^{r_{2}\dagger}\gamma^{r_{1}%
}t\right)  \right\vert _{t^{N/2}}\label{GCM2}\\
\left\langle \gamma^{r_{2}}\right\vert c_{a}^{\dagger}c_{b}\left\vert
\gamma^{r_{1}}\right\rangle  &  =\,\left.  \left(  \frac{\gamma^{r_{1}}%
\gamma^{r_{2}\dagger}t}{1+\gamma^{r_{1}}\gamma^{r_{2}\dagger}t}\right)
_{ba}\mathrm{pf}\left(  1+\gamma^{r_{2}\dagger}\gamma^{r_{1}}t\right)
\right\vert _{t^{N/2}}\label{GCM3}\\
\left\langle \gamma^{r_{2}}\right\vert c_{a}^{\dagger}c_{b}^{\dagger}%
c_{d}c_{c}\left\vert \gamma^{r_{1}}\right\rangle  &  =\left(  2\left[
\frac{\gamma^{r_{1}}\gamma^{r_{2}\dagger}t}{1+\gamma^{r_{1}}\gamma
^{r_{2}\dagger}t}\right]  _{ca}\left[  \frac{\gamma^{r_{1}}\gamma
^{r_{2}\dagger}t}{1+\gamma^{r_{1}}\gamma^{r_{2}\dagger}t}\right]  _{db}\right.
\nonumber\\
&  \left.  \left.  +t\left[  \frac{1}{1+\gamma^{r_{1}}\gamma^{r_{2}\dagger}%
t}\gamma^{r_{1}}\right]  _{cd}\left[  \gamma^{r_{2}\dagger}\frac{1}%
{1+\gamma^{r_{1}}\gamma^{r_{2}\dagger}t}\right]  _{ba}\right)  \mathrm{pf}%
\left(  1+\gamma^{r_{2}\dagger}\gamma^{r_{1}}t\right)  \right\vert _{t^{N/2}},
\label{GCM5}%
\end{align}
where $\gamma^{\dagger}$ corresponds to the Hermitian conjugate of $\gamma$
and pf denotes the Fredholm Pfaffian which is the square root of the Fredholm
determinant%
\begin{equation}
\mathrm{pf}\left(  1+\gamma^{r_{2}\dagger}\gamma^{r_{1}}t\right)
=\sqrt{\mathrm{\det}\left(  1+\gamma^{r_{2}\dagger}\gamma^{r_{1}}t\right)  }.
\end{equation}
It was shown in Ref.\cite{Neergard1983} that $\mathrm{pf}\left(
1+\gamma^{r_{2}\dagger}\gamma^{r_{1}}t\right)  $ is a polynomial of degree
$M/2$ since roots of the characteristic polynomial of a product of two
antisymmtric matrices are pairwise degenerated.

The total energy can be obtained from the matrix element of the Hamiltonian
($\mathcal{H}$)%
\begin{equation}
\left\langle \gamma^{r_{2}}\right\vert \mathcal{H}\left\vert \gamma^{r_{1}%
}\right\rangle =\sum_{ab}h_{ab}\left\langle \gamma^{r_{2}}\right\vert
c_{b}^{\dagger}c_{a}\left\vert \gamma^{r_{1}}\right\rangle +\frac{1}{2}%
\sum_{abcd}V_{abcd}\left\langle \gamma^{r_{2}}\right\vert c_{a}^{\dagger}%
c_{b}^{\dagger}c_{d}c_{c}\left\vert \gamma^{r_{1}}\right\rangle ,
\end{equation}
where $h$ corresponds to the sum of the kinetic energy and the external
potential and $V$ is the Coulomb repulsion.\ Contrary to standard GCM schemes,
we explicitly obtain the coefficient of $t^{N/2}$ in Eqs.(\ref{GCM1}%
-\ref{GCM5}) instead of using the particle projection method\cite{Onishi1966,
Neergard1983}. A similar projection method\cite{Onishi1966, Neergard1983} can
be applied to the HFB state to adapt the wavefunction to the spatial and spin
symmetries, which is necessary to obtain accurate ground state wavefunctions,
but we do not use it in the present study to compare ESTD and full-CI without
the symmetry adaptation. We also explicitly obtained the derivatives of the
energy with respect to $\gamma $, with which to optimize the
parameters $U_{ia}^{r}$ using the Broyden-Fletcher-Goldfarb-Shanno algorithm
(BFGS)\cite{Newton1, Newton2, Newton3, Newton4, Newton5}. We emphasize that
ESTD allows the handling of non-orthogonal AOs by replacing $\gamma_{ab}^{r}$
by $\left(  s^{1/2}\gamma^{r}s^{1/2}\right)  _{ab}$ where $s$ is the overlap
matrix of AOs, although orthogonality is assumed for simplicity in the present formulation.

\bigskip When using general complex matrices for $U_{ia}^{r}$ in practical
numerical works, the computation is sometimes impossible to carry on because
of the loss of significance in the floating-point arithmetic. To avoid this
problem, we make the geminal complex unitary as well as antisymmetric so that
all the eigenvalues are equal to unity in the absolute values although the convergence
speed is slightly affected. Note that the geminal can be made unitary by the
following redefinition of $U_{ia}^{r}$: noticing that the second order
permutation tensor $\epsilon$ is real antisymmetric, we can apply the
canonical tranformation\cite{BM} by which%
\begin{equation}
\epsilon=\bar{U}^{T}\left(
\begin{array}
[c]{cc}%
0 & \Lambda\\
-\Lambda & 0
\end{array}
\right)  \bar{U}=\bar{U}^{T}\left(
\begin{array}
[c]{cc}%
0 & \Lambda^{1/2}\\
\Lambda^{1/2} & 0
\end{array}
\right)  \left(
\begin{array}
[c]{cc}%
0 & I\\
-I & 0
\end{array}
\right)  \left(
\begin{array}
[c]{cc}%
0 & \Lambda^{1/2}\\
\Lambda^{1/2} & 0
\end{array}
\right)  \bar{U},
\end{equation}
where $\Lambda$ and $I$ are, respectively, real diagonal and identity matrix,
$\bar{U}$ is a real orthogonal matrix, and the superscript $T$ means taking
the matrix transpose: then the geminal (Eq. (\ref{gamma})) becomes%
\begin{equation}
\gamma^{r}=\left(  U^{r}\right)  ^{T}\left(
\begin{array}
[c]{cc}%
0 & I\\
-I & 0
\end{array}
\right)  U^{r}%
\end{equation}
where $U^{r}$ has been redefined as%
\begin{equation}
U^{r}\rightarrow\left(
\begin{array}
[c]{cc}%
0 & \Lambda^{1/2}\\
\Lambda^{1/2} & 0
\end{array}
\right)  \bar{U}U^{r}.
\end{equation}
Since $U^{r}$ was introduced in Eq.(\ref{extended}) as a general matrix to
be determined variationally, this redefinition does not affect the resulting
formulas. In this way the geminal becomes complex unitary when restricting
$U_{ia}^{r}$ to be complex unitary. We call such restricted ESTD the
antisymmetrized unitary geminal powers (AUGP) hereafter. The variation of the
unitary matrix can be done using the Cayley transform
\begin{equation}
U=(1+X)^{-1}(1-X)
\end{equation}
with a skew-Hermitian matrix $X$.

\section{Results}

We have applied ESTD and AUGP to the water molecule and half-filled Hubbard
tetramers and compared the total energy with the STD-CI and the full-CI
calculations. In the water molecule, we set the geometry condition to
\begin{equation}
O=(0,0,0),\,H=(-1.809,0,0),\,(0.453549,1.751221,0)
\end{equation}
in atomic units. The calculation was done using the basis set STO-3G. The
minimal basis set is sufficient for comparison between the different schemes,
although larger basis sets will be required for highly quantitative
calculations. The Hubbard tetramer has a tetrahedral geometry, and all the
sites are connected by the transfer integral of unity. The Hubbard $U$ was
taken as 1, 10$^{\text{2}}$, and 10$^{\text{4}}$.

\begin{figure}[ptb]
\begin{center}
\includegraphics[
height=2.0in,
width=3.0in
]
{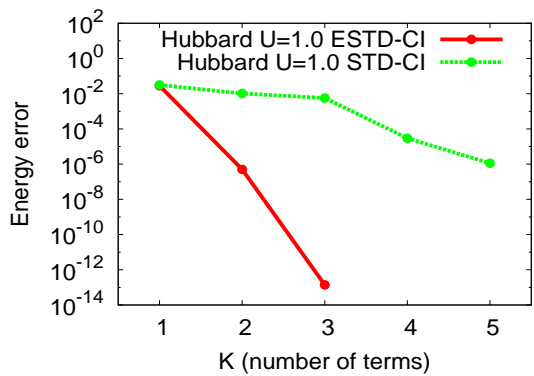}
\end{center}
\caption{Energy error of 4 sites $U$=1.0 Hubbard model for ESTD (solid line with
circle) and STD-CI (broken line with circle).}%
\label{hubbard1}%
\end{figure}

\begin{figure}[ptb]
\begin{center}
\includegraphics[
height=2.0in,
width=3.0in
]
{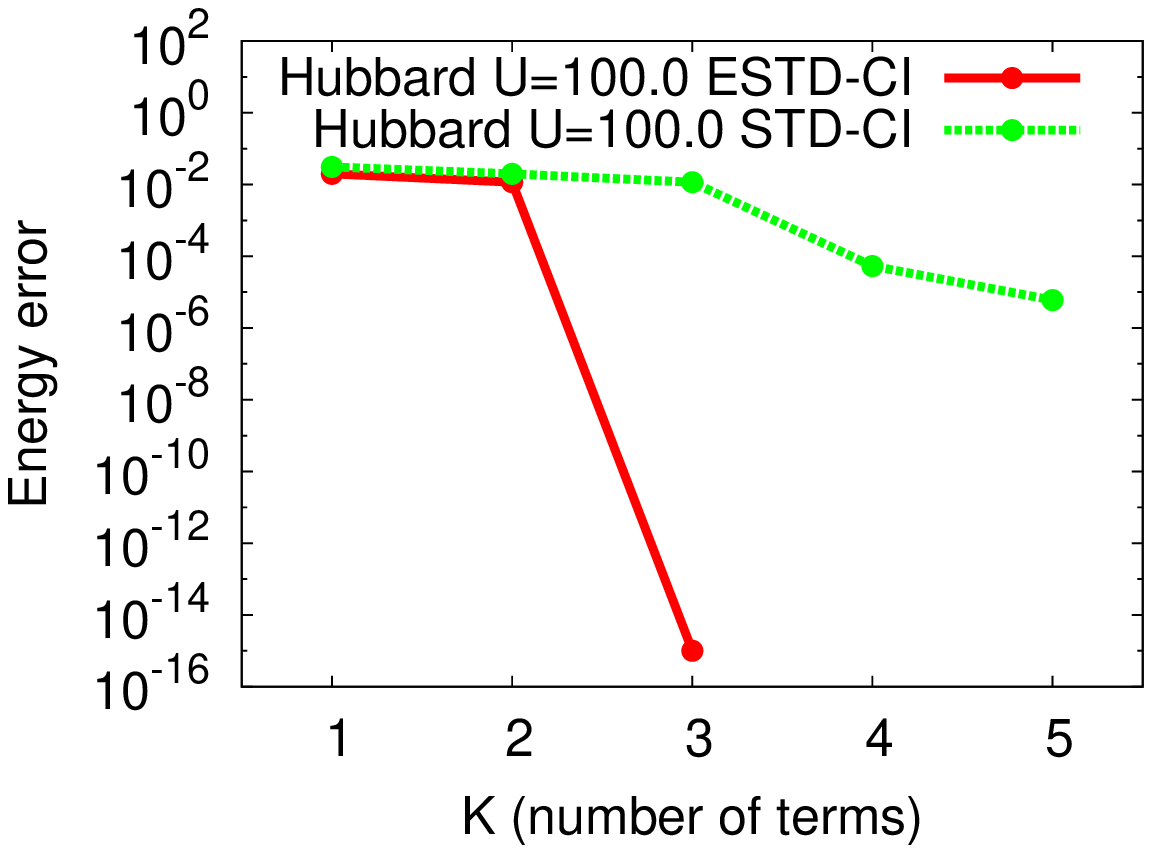}
\end{center}
\caption{Energy error of 4 sites $U$=100.0 Hubbard model for ESTD (solid line
with circle) and STD-CI (broken line with circle).}%
\label{hubbard100}%
\end{figure}

\begin{figure}[ptb]
\begin{center}
\includegraphics[
height=2.0in,
width=3.0in
]
{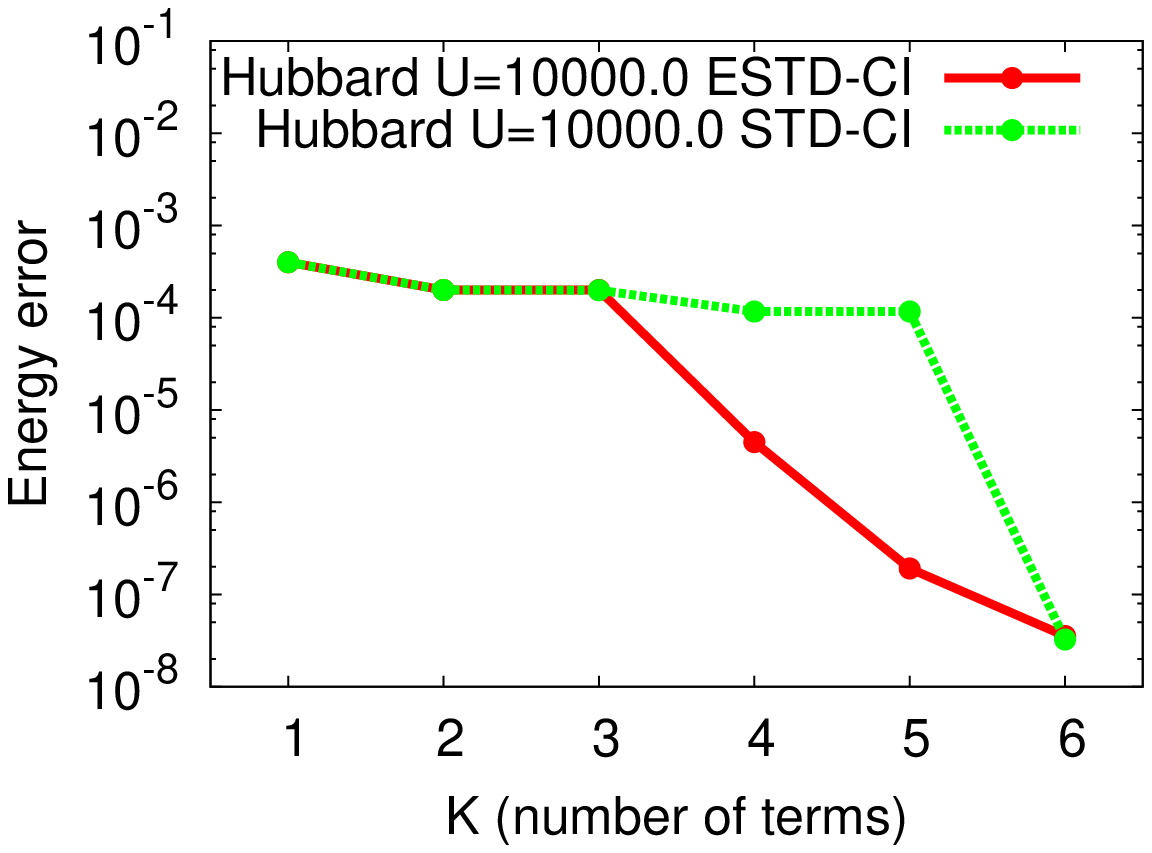}
\end{center}
\caption{Energy error of 4 sites $U$=10000.0 Hubbard model for ESTD (solid line
with circle) and STD-CI (broken line with circle).}%
\label{hubbard10000}%
\end{figure}

Figures \ref{hubbard1}, \ref{hubbard100} and \ref{hubbard10000} show the total
energy referred to the full-CI calculation, or the residual error vs. the
tensor rank of the decomposition. The convergence is much faster for ESTD than
for STD-CI. For larger $U$ cases, the residual error is initially almost
insensitive to the rank and the value amounts to 1-10 when rescaled in unit of
$U$. Then the residual error drops suddenly when $K$ is larger than 2 for
$U=10^{2}$ and 3 for $U=10^{4}$. This might indicate that the
anti-ferromagnetic (AF) ground state cannot be described even at a qualitative
level when the number of parameters is unreasonably small. For $U=10^{4}$, the
total energy is almost the same among ESTD, STD-CI, and full-CI when the rank
is 6. We postulate that the parameter set for STD-CI happens to become
suitable for describing the AF limit only at $K\geq6$.

Table \ref{tb:hubbard100AUGP} shows detailed comparison of the total energy
calculated by AUGP and full-CI for the Hubbard tetramer with $U=10^{2}$. The
calculation was done using double precision with $K=3$ . The residual error is
$2.2\times10^{-11}$ while the error was around $10^{-15}$ for ESTD (Fig.
(\ref{hubbard100})); in either scheme the error is close to the double precision
limit. \begin{table}[tbh]
\begin{center}%
\begin{tabular}
[c]{|l|l|}\hline
Method & Total energy\\\hline
AUGP & $-0.119880248924950$\\
Exact & $-0.119880248946222$\\\hline
\end{tabular}
\end{center}
\caption{Total energy of the Hubbard tetramer with $U=100$ obtained by the
AUGP with $K$ = 3. The result obtained by full-CI is also shown.}%
\label{tb:hubbard100AUGP}%
\end{table}

\begin{figure}[ptb]
\begin{center}
\includegraphics[
height=2.0in,
width=3.0in
]{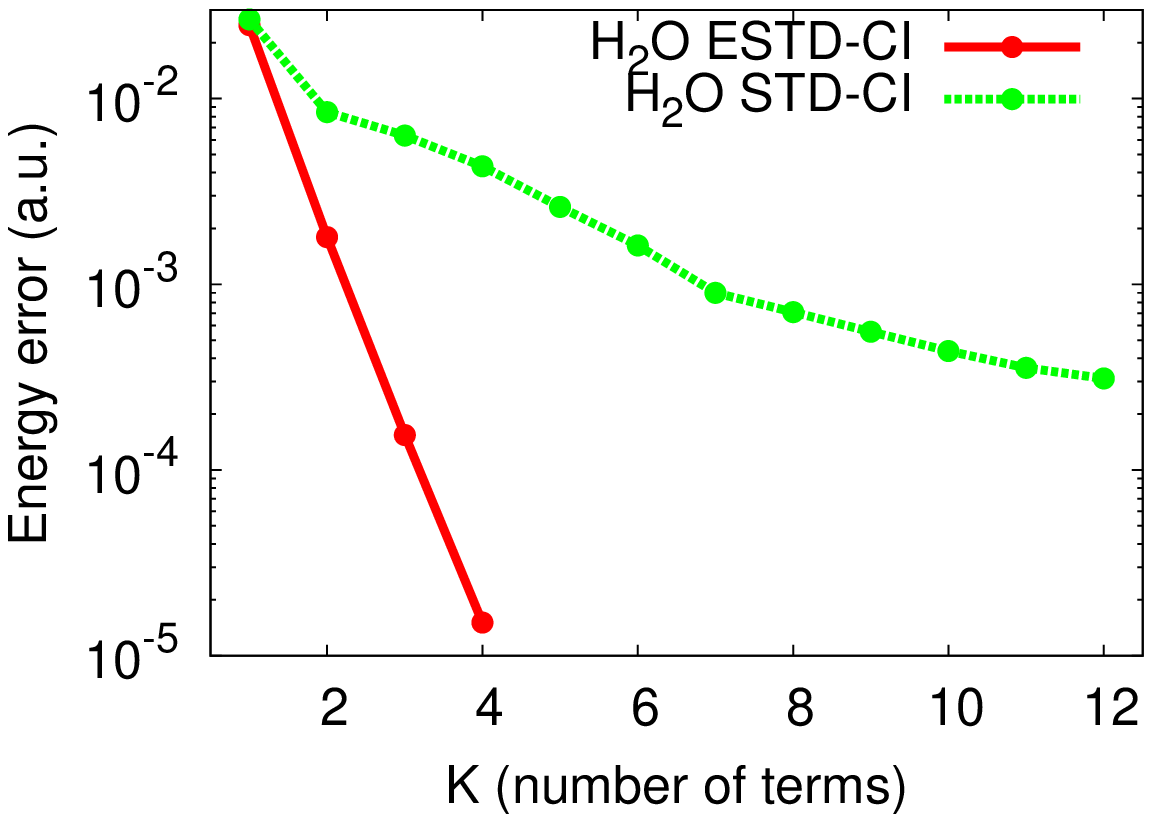}
\end{center}
\caption{Residual error of the total energy of the H$_{2}$O molecule obtained by
ESTD (solid line with circle) and STD-CI (broken line with circle).}%
\label{figh2o}%
\end{figure}

\begin{figure}[ptb]
\begin{center}
\includegraphics[
height=2.0in,
width=3.0in
]
{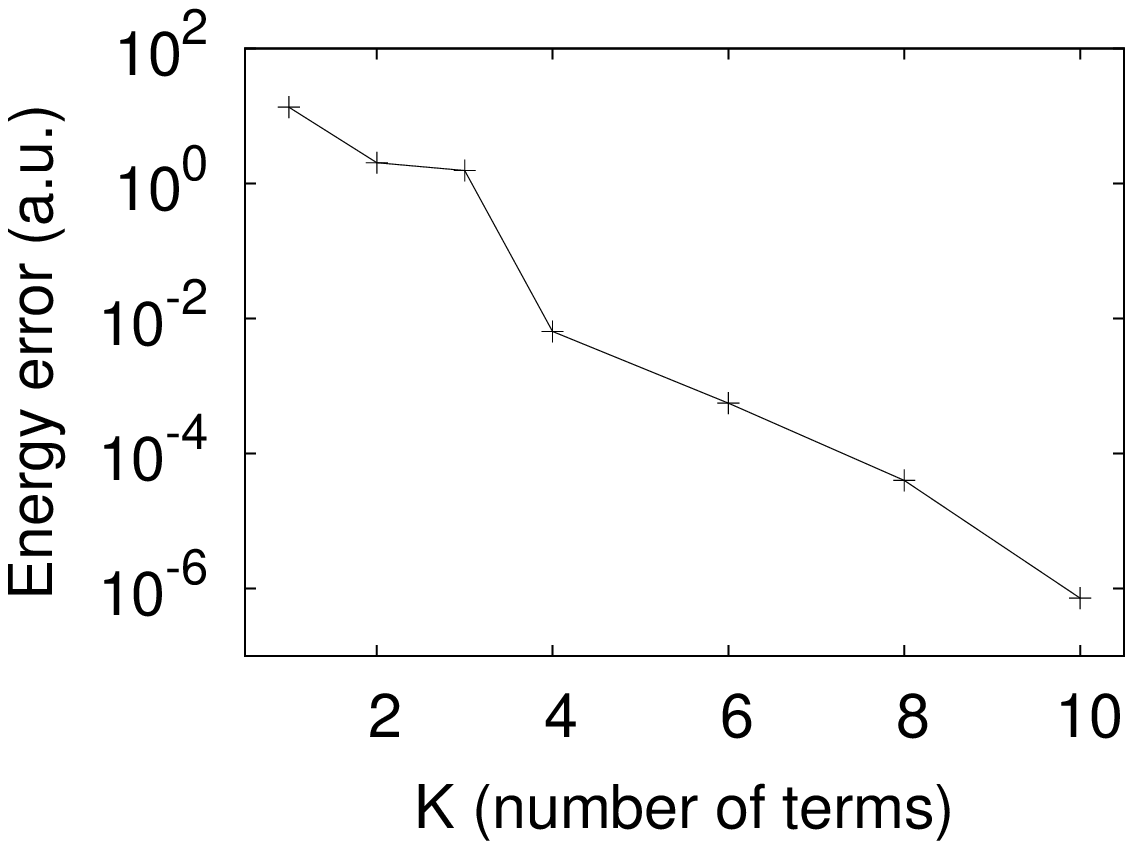}
\end{center}
\caption{Residual error of the total energy of H$_{2}$O obtained by AUGP}%
\label{h2oerror}%
\end{figure}

Figure \ref{figh2o} shows the residual error of the total energy of the
H$_{2}$O molecule. ESTD shows strikingly faster convergence than STD-CI. In
the ESTD and STD-CI calculations, we used quadruple precision since serious
loss of significance occurred with double precision. Because of this problem,
we could not easily obtain accurate results using $K$ larger than 4 for ESTD,
and thus we could not reduce the residual error below $10^{-5}$ Hartree.
Figure \ref{h2oerror} and Table \ref{tb:h2o_augp} show the result obtained by
using AUGP with double precision. In obtaining the residual error, we used the
total energy from full-CI calculation as the reference, but the result is
almost the same when using the complete active space self-consistent field
(CASSCF) calculation performed using GAUSSIAN09 package\cite{Gaussian}. The
figure shows that the residual energy is reduced almost linearly in log-scale
with respect to $K$, indicating an exponential convergence. The table shows
that the error is $6.4\times10^{-3}$ Hartree with $K=4$ and is reduced to
$7.2\times10^{-7}$ Hartree with $K=10$. Although AUGP yields slower
convergence than ESTD, the AUGP is very stable. It is not difficult to obtain the
optimal parameters $U_{ia}^{r}$ even when the calculation is started from
random initial values.

\begin{table}[tbh]
\begin{center}%
\begin{tabular}
[c]{|l|l|}\hline
Method & Total energy\\\hline
AUGP, $K=1$ & -61.508355\\
AUGP, $K=2$ & -72.985876\\
AUGP, $K=3$ & -73.457483\\
AUGP, $K=4$ & -75.006008\\
AUGP, $K=6$ & -75.011870\\
AUGP, $K=8$ & -75.012385975\\
AUGP, $K=10$ & -75.012425100\\\hline
Exact (ours) & -75.012425818\\
Exact (Gaussian) & -75.012425839\\\hline
\end{tabular}
\end{center}
\caption{Total energy (in unit of Hartree) of H$_{2}$O obtained by AUGP. For
comparison, the full-CI calculation was done using our own code and the CASSCF
calculation using the Gaussian09 package.}%
\label{tb:h2o_augp}%
\end{table}

\section{Conclusion}

To efficiently represent the CI wavefunction, we have applied the canonical
decomposition algorithm to the symmetric part of the CI coefficient. Therein
the molecular orbitals are fully optimized, without imposing the
orthonormality condition, differently for different terms in the decomposition
series. The computational scheme, which we call the extended symmetric tensor
decomposition (ESTD), is equivalent to the linear combination of the
Hartree-Fock-Bogoliubov states or the linear combination of the
antisymmetrized geminal power (AGP). By this, we can rearrange the full-CI
series into the canonical decomposition series (STD series). By applying ESTD
to the water molecule and Hubbard tetramers, we found that the total energy
rapidly converges well within ten terms ($K=10$) for the STD series. The ESTD
calculation was found to be numerically unstable when the number of electrons
is increased as large as 10 because of the loss of significance in the
floating-point arithmetic. This problem could be avoided by restricting the MO
coefficients to be complex unitary although the convergence speed with respect
to tensor rank $K$ was slightly affected. The computational cost of ESTD
scales as $K^{2}M^{5}$ where $M$ is the number of MOs. The result suggests
that the AGP-based scheme is a promising computational tool for quantum
systems. In this context, it will be an important target of future study to
clarify how $K$ depends on the complexity and the size of the system. Our
calculation was done without parallelization, but an acceleration by a factor of
$K^{2}$ can be expected because of the almost independent nature of the
computation. Further acceleration is expected by applying the tensor
decomposition scheme to the two-electron integrals as done in Ref.\cite{MP2CP}%
; such technique may possibly reduce the scaling to $K^{2}M^{4}$.

\bigskip

\section{Acknowledgements}

We acknowledge the Strategic Programs for Innovative Research by the Ministry
of Education, Culture, Sports, Science and Technology of Japan and the
Computational Materials Science Initiative for the financial support during
our research. Some of the calculations were performed in the supercomputer
facility at the Institute for Solid State Physics, the University of Tokyo. S.
K. is supported by Grant-in-Aid for Research Activity Start-up (grant number
25887017) by the Japan Society for the Promotion of Science.


\begin{thebibliography}{99}                                                                                               %


\bibitem {shavitt}I. Shavitt, Mol. Phys. 94, 3 (1998).

\bibitem {foulkes}W. M. C. Foulkes, L. Mitas, R. J. Needs and G. Rajagopal,
Rev. Mod. Phys. 73, 33 (2001).

\bibitem {springer}D. C. Cabra, A. Honecker and P. Pujol, \textit{Modern
theories of Many-Particle Systems in Condensed Matter Physics}
(Springer-Verlag, Berlin, Heidelberg, 2012).

\bibitem {rank}In this paper, we use \textquotedblleft order\textquotedblright%
\ for the number of indices required to represent a tensor such that a matrix
is a tensor of order $2$ and use \textquotedblleft rank\textquotedblright\ for
the number of terms required for the tensor decomposition.

\bibitem {LMO}T. Neuhauser, M. von Armin and S. D. Peyermhoff, Theor. Chim.
Acta. 83, 123 (1992).

\bibitem {Goto}H. Goto, M. Kojo, A. Sasaki and K. Hirose, Nanoscale Res.
Lett. 8, 200 (2013).

\bibitem {bartlett}R. J. Bartlett and M. Musia\l , Rev. Mod. Phys. 79, 291 (2007).

\bibitem {uemura2012}W. Uemura and O. Sugino, Phys. Rev. Lett. 109, 253001 (2012).

\bibitem {CANDECOMP}R. A. Harshman, UCLA Working Papers in Phonetics, 16, 1 (1970).

\bibitem {PARAFAC}J. D. Carroll and J.J. Chang, Psychometrika 35, 283 (1970).

\bibitem {white1992}S. R. White, Phys. Rev. Lett. 69, 2863 (1992).

\bibitem {white1993}S. R. White, Phys. Rev. B 48, 10345 (1993).

\bibitem {PEPS}F. Verstraete and J. Cirac, arXiv:cond-mat/0407066
(unpublished); V. Murg, F. Verstraete and J. I. Cirac, Phys. Rev. A 75,
033605 (2007).

\bibitem {chan2002}G. K.-L. Chan and M. Head-Gordon, J. Chem. Phys. 116, 4462 (2002).

\bibitem {HFB}N. N. Bogoliubov, Sov. Phys. Usp. 2, 236 (1959), M. Baranger,
1962 Cargese lectures in theoretical physics (Benjamin, New York, 1963).

\bibitem {coleman1}A. J. Coleman, Rev. Mod. Phys. 35, 668 (1963).

\bibitem {coleman2}A. J. Coleman, J. Math. Phys. 6, 1425 (1965).

\bibitem {Ortiz1981}J. V. Ortiz, B. Weiner and Y. Ohn, Int. J. Quantum Chem. 15,
113 (1981).

\bibitem {Straroverov2002}V. N. Starovecrov and G. E. Scuseria, J. Chem. Phys.
117, 11107 (2002).

\bibitem {Scuseria2011}G. E. Scuseria, C. A. Jimenez-Hoyos, T. M. Henderson,
K. Samanta and J. K. Ellis, J. Chem. Phys. 135, 124108 (2011).

\bibitem {Kobayashi2014}M. Kobayashi, J. Chem. Phys. 140, 084135 (2014).

\bibitem {weiner}B. Weiner and J. V. Ortiz, J. Chem. Phys. 117, 5135 (2002).

\bibitem {eric}E. Neuscamman, Phys. Rev. Lett. 109, 203001 (2012).

\bibitem {sorella}A. Zen, E. Coccia, Y. Luo, S. Sorella and L. Guidoni, J.
Chem. Theory Comput. 10 (3), 1048 (2014).

\bibitem {Tahara2008}D. Taraha and M. Imada, J. Phys. Soc. Jpn. 77, 114701 (2008).

\bibitem {bajdich2006}M. Bajdich, L. Mitas, G. Drobny, L. K. Wagner and K. E.
Schmidt, Phys. Rev. Lett. 96, 130201 (2006).

\bibitem {bajdich2008}M. Bajdich, L. Mitas, L. K. Wagner and K. E. Schmidt,
Phys. Rev. B 77, 115112 (2008).

\bibitem {GCM}J. J. Griffin and J. A. Wheeler, Phys. Rev. 108, 311 (1957).

\bibitem {Onishi1966}N. Onishi and S. Yoshida, Nucl. Phys. 80, 367 (1966).

\bibitem {doba}J. Dobaczewski, Phys. Rev. C 62, 017301 (2000).

\bibitem {Mizusaki2012}T. Mizusaki and M. Oi, Physics Letters B, 715, 219 (2012).

\bibitem {Neergard1983}K. Neerg\aa rd and W. W\"{u}st, Nucl. Phys. A 402, 311 (1983).

\bibitem {Newton1}C. G. Broyden, J. lnst. Maths. Appl. 6, 76 (1970).

\bibitem {Newton2}R. Fletcher, Comp. J. 13, 317 (1970).

\bibitem {Newton3}D. Goldfarb, Math. Comp. 24, 23 (1970).

\bibitem {Newton4}D. F. Shanno, Math Comp. 24, 647 (1970).

\bibitem {Newton5}R. Fletcher, in \textit{Practical methods of optimization},
Vol. 1 (Wiley, New York, 1980).

\bibitem {BM}C. Bloch and A. Messiah, Nucl. Phys. 39, 95 (1962).

\bibitem {Gaussian}Gaussian 09, M. J. Frisch, G. W. Trucks, H. B. Schlegel, G.
E. Scuseria, M. A. Robb, J. R. Cheeseman, G. Scalmani, V. Barone, B. Mennucci,
G. A. Petersson, H. Nakatsuji, M. Caricato, X. Li, H. P. Hratchian, A. F.
Izmaylov, J. Bloino, G. Zheng, J. L. Sonnenberg, M. Hada, M. Ehara, K. Toyota,
R. Fukuda, J. Hasegawa, M. Ishida, T. Nakajima, Y. Honda, O. Kitao, H. Nakai,
T. Vreven, J. A. Montgomery, Jr., J. E. Peralta, F. Ogliaro, M. Bearpark, J.
J. Heyd, E. Brothers, K. N. Kudin, V. N. Staroverov, R. Kobayashi, J. Normand,
K. Raghavachari, A. Rendell, J. C. Burant, S. S. Iyengar, J. Tomasi, M. Cossi,
N. Rega, J. M. Millam, M. Klene, J. E. Knox, J. B. Cross, V. Bakken, C. Adamo,
J. Jaramillo, R. Gomperts, R. E. Stratmann, O. Yazyev, A. J. Austin, R. Cammi,
C. Pomelli, J. W. Ochterski, R. L. Martin, K. Morokuma, V. G. Zakrzewski, G.
A. Voth, P. Salvador, J. J. Dannenberg, S. Dapprich, A. D. Daniels, \"{O}.
Farkas, J. B. Foresman, J. V. Ortiz, J. Cioslowski and D. J. Fox, Gaussian,
Inc., Wallingford CT (2009).

\bibitem {MP2CP}U. Benedikt, A. A. Auer, M. Espig and W. Hackbusch, J. Chem.
Phys. 134, 054118 (2011).
\end{thebibliography}
\end{document}